\documentstyle[12pt]{article}
\begin{document}
\baselineskip 15pt plus 2pt
\newcommand{\be}{\begin{equation}}
\newcommand{\ee}{\end{equation}}

\hspace{10.5cm}
BRX-TH-357
\vspace{1cm}
\begin{center}
\begin{bf}
\begin{Large}
Singularity Free Quasi-Classical \\
Schwarzschild Space-Times \\
\end{Large}
\end{bf}
\vspace{1cm}
Yoav Peleg\footnote{This work is supported by the NSF grant
PHY 93-15811}\\
\vspace{0.3cm}
Physics Department, Brandeis University, Waltham, MA 02254\\
\vspace{0.2cm}
{\em Peleg@Brandeis}\\
\vspace{1.5cm}
\begin{large}
ABSTRACT\\
\end{large}
\end{center}
\vspace{0.2cm}
Using canonical (Schr\"{o}dinder) quantization of spherically
symmetric gravitational dust systems, we find the
quasi-classical (coherent) state, $|\alpha^{(s)}>$,
that corresponds to
the classical Schwarzschild solution. We calculate
the ``quasi-classical Schwarzschild metric", which is
the expectation value of the quantized metric
in this quasi-classical state, $g^{(\alpha)}_{\mu \nu} =
<\alpha^{(s)}| \hat{g}_{\mu \nu} |\alpha^{(s)}>$.
Depending on the
quantization scheme that we use, we study three different
quasi-classical geometries, for all of which
$g^{(\alpha)}_{\mu \nu}$ turns out to
be singularity free. Their maximal extensions are
complete manifolds with no singularities, describing
a tower of asymptotically flat universes connected
through Planck size wormholes.

\newpage

Spherically symmetric dust matter in 4D Einstein gravity
is an integrable system [1].
One can fix the (diffeomorphism) gauge completely,
and get an infinite set of independent (non propagating)
degrees of freedom [2]. Though very simple, this dust
system is rich enough to have some physically interesting
(classical) solutions. Two of these are the homogeneous
dust-filled Robertson-Walker universe and the
Oppenheimer-Snyder collapsing star. The first gives us
a simple model of the big bang, and the second is
a simple model of black hole formation.
Because of the simplicity of this dust system, it can be
easily quantized [2,3]. In this letter we study some of the
effects of this quantization. Especially, we find the
quasi-classical metric, and study its structure.

Following Lund [2], we first fix the gauge
classically, and then quantized the reduced system.
One can fix the gauge by choosing comoving coordinates
in which the metric has the form
   \be
ds^{2} = -d\tau^{2} + e^{2\mu}d {\cal R}^{2} + e^{2\lambda}d\Omega_{2}^{2}
   \ee
where $\mu$ and $\lambda$ are functions of $\tau$ and ${\cal R}$, and
$d\Omega_{2}^{2}$ is the volume element in $S^{2}$. Explicitly, we choose
$\tau$ and ${\cal R}$ such that
   \begin{eqnarray}
U_{\nu} &=& - \nabla_{\nu} \tau \nonumber \\
{\cal R} &=& \lambda' e^{\lambda - \mu}
   \end{eqnarray}
where $U_{\nu}$ are the 4-velocities of the dust particles,
and the primes denote ${\cal R}$-differentiation. This gauge
fixing corresponds to $N_{\perp}=1$ , $N^{i}=0$ , and it is complete
[1,2].The resultant reduced Hamiltonian [2,3] is
   \be
H_{red} = \int_{0}^{\rho_{s}} {\cal H}(\rho) =
\int_{0}^{\rho_{s}} \left( \frac{4 \rho \sqrt{1-\rho^{2}}}{3\pi}
\frac{P_{Y}^{2}}{Y} +
\frac{12 \pi \rho}{\sqrt{1 - \rho^{2}}} Y \right) d\rho
   \ee
where $\rho = \sqrt{1 - {\cal R}^{2}}$ , $\rho_{s}$ is the
surface of the dust star,
$Y(\tau,\rho)$ is the (reduced) field variable, $ Y= e^{\lambda} $,
and $P_{Y} = \delta L / \delta \dot{Y} = 3\pi Y \dot{Y} / (8 \rho
\sqrt{1 - \rho^{2}}) $.
In the $(\tau , \rho , \theta , \phi )$ coordinates, the
metric (1) is
  \be
ds^{2} = - d\tau^{2} + (\partial_{\rho} Y)^{2} \frac{d\rho^{2}}{1
- \rho^{2}} + Y^{2} d \Omega_{2}^{2}
  \ee

The Hamiltonian (3) is time independent, namely
${\cal H}(\rho) = E(\rho) $,
where $E(\rho)$ is a function of $\rho$ only. Classically, $E(\rho)$
is determined by the Cauchy data, $Y(\tau=0)$ and $\dot{Y}(\tau=0)$.
A most important feature of (3) is that there
are no $\rho$-derivatives in it. This means that the infinite number
of degrees of freedom are decoupled. Each can be described as
an harmonic oscillator [3]. To see this it is more convenient
to use a different gauge. Consider the gauge $N_{\perp}=Y/\rho$
and $N_{\rho}=-\int^{\eta} \partial_{\rho} (Y/\rho) d\eta'$, which
corresponds to choosing the time coordinate $\eta$ instead of $\tau$,
where
  \be
\tau = \int^{\eta} \frac{Y}{\rho} d\eta'
  \ee
The metric in this gauge is
  \begin{eqnarray}
ds^{2} &=& - \frac{Y^{2}}{\rho^{2}} d\eta^{2} + \left[
\frac{ (\partial_{\rho} Y)_{\tau}^{2} }{1-\rho^{2}} -
{\left( \int^{\eta} \partial_{\rho} (Y/\rho) d\eta' \right)}^{2}
\right] d\rho^{2}  \nonumber \\
& & - \left[ \int^{\eta} \partial_{\rho} (Y/\rho)^{2} d\eta'
\right] d\eta d\rho + Y^{2} d\Omega_{2}^{2}
  \end{eqnarray}
where $(\partial_{\rho} )_{\tau}$ denotes $\rho$ derivative in which
$\tau$ is held fixed, while $\partial_{\rho}$ (with no $\tau$ index)
denotes a $\rho$ derivative where $\eta$ is fixed.
Redefinition of the field
  \be
X(\eta,\rho) \equiv Y(\eta ,\rho ) - \frac{1}{3\pi} \sqrt{1-\rho^{2}}
E(\rho) ~,
  \ee
implies that the following modified Hamiltonian, $\tilde{H}$,
is $\eta$-independent,
  \be
\tilde{H} = \int_{0}^{\rho_{s}}
\left( \frac{\sqrt{1-\rho^{2}}}{6\pi}
P_{X}^{2} + \frac{3\pi}{2\sqrt{1 - \rho^{2}}} X^{2} \right) d\rho =
\int_{0}^{\rho_{s}} \frac{\sqrt{1-\rho^{2}}}{6\pi} E^{2}(\rho) d\rho
  \ee
where $P_{X} = 3\pi (8 \sqrt{1-\rho^{2}})^{-1} (\partial_{\eta}
X)^{2}$. Clearly, the system described by (8) is an
infinite set of (decoupled) harmonic oscillators\footnote{To see
that, one can discretize $\rho$ by defining
$ \rho_{k} = \frac{k}{N} \rho_{s} $, $ k = 1,2,...,N $.
Eq. (8) is then replaced by a set of $N$ (one dimensional)
harmonic oscillators, satisfying
  \begin{eqnarray*}
\tilde{{\cal H}}_{k} = \alpha_{k} P^{2}_{k}(\eta) + \beta_{k}
X^{2}_{k}(\eta)= \tilde{E}_{k}
  \end{eqnarray*}
where $X_{k}(\eta)=X(\eta,\rho_{k})$, $P_{k}(\eta)=P(\eta,\rho_{k})$,
$\tilde{E}_{k} = \tilde{E}(\rho_{k})$, and
$ \alpha_{k} = \frac{ \sqrt{1-\rho_{k}^{2}}}{6\pi} $,
$ \beta_{k} = \frac{3\pi}{2 \sqrt{1 - \rho^{2}_{k}}} $.
In the continuum limit, $N \rightarrow \infty$, this is an
infinite set.}.

Considering classical solutions,we choose $\tau=0$ (or
equivalently $\eta=0$) such that\footnote{
Of course this is possible only for a certain class of solutions,
but we are interested only in this class.}
$P(\eta=0,\rho)=0$. In this case, the
general classical solution of (8) (or (3)) is
  \be
X(\eta,\rho) = X_{0}(\rho) \mbox{cos}(\eta)
  \ee
where $X_{0}(\rho) = X(\eta=0,\rho)$.
For a given initial data, $X_{0}(\rho)$, or equivalently for a
given initial dust density $\rho_{dust}(\eta=0, \rho)$, we
have a unique solution (9).

There are two interesting (classical) solutions:
the homogeneous density solution, and the zero density one.
For the homogeneous density case, we have
  \be
X_{0} \equiv X_{0}^{(hom)}(\rho) = \frac{1}{2} a_{0} \rho
{}~~~\mbox{and}~~~ Y^{(hom)}(\tau,\rho) = \frac{1}{2}a_{0} \rho
(1 + \mbox{cos}\eta )
  \ee
where $a_{0}$ is a constant and $\tau = \frac{1}{2}a_{0}(\eta
+ \mbox{sin}(\eta) ) $. The metric (4) is now
  \be
ds^{2} = -d\tau^{2} + {\left( \frac{1}{2}a_{0}(1 + \mbox{cos}(\eta) )
\right)}^{2} (\frac{d\rho^{2}}{1-\rho^{2}}
+ \rho^{2} d\Omega_{2}^{2} ) ~,
  \ee
which is the well known cycloidal solution for
a homogeneous dust-filled universe.

For the zero density solution we have
  \be
X_{0} \equiv X_{0}^{(Sch)}(\rho) = \frac{M}{\rho^{2}}
  \ee
or
  \begin{eqnarray}
Y^{(Sch)}(\tau,\rho) &=& \frac{M}{\rho^{2}} (1 + \mbox{cos}(\eta) )
\nonumber \\
\tau &=& \frac{M}{\rho^{3}} (\eta + \mbox{sin}(\eta))
  \end{eqnarray}
Here $M$ is a constant, and we will see that it is indeed the
Schwarzschild mass.
The uniqueness theorem, tells us that the only spherically
symmetric zero density solution is the Schwarzschild one.
To see that (13) is indeed the Schwarzschild solution, we
write it in Schwarzschild coordinates $(t,r,\theta,\phi)$.
The metric in the $(\tau,\rho)$ coordinate is (4) with (13),
from which we see that the Schwarzschild
radial coordinate, $r$, should be defined such that
  \be
r = Y^{(Sch)}(\eta,\rho)
  \ee
Define the Schwarzschild time, $t$, such that $g_{tt} = -1/g_{rr}$,
to get the familiar form
  \be
(ds^{2})_{Sch} =  - \left( 1 - \frac{2 M}{r} \right) dt^{2} +
\frac{dr^{2}}{1 - 2M/r} + r^{2} d\Omega_{2}^{2}
  \ee

An interesting feature of these two solutions, is that they
can be matched continuously at the surface of the dust star,
$\rho = \rho_{s}$ [4,5]. From (10) and (13) we see that
the matching condition is\footnote{Together
with (14), that says $r_{s}(\tau) = a(\tau) \rho_{s}$.}
  \be
2M = a_{0} \rho_{s}^{3}
  \ee
This is the Oppenheimer-Snyder (OS) collapsing star [4].

The coordinates $(\tau,\rho,\theta,\phi)$ are the Novikov
coordinates, where our $\rho$ is related to the original
Novikov coordinate, $R^{*}$, by $\rho^{2} = (R^{*} + 1)^{-2}$.
The range of $\rho$ is
$0 \leq \rho \leq \rho_{s} < 1$.In the OS solution,
$\rho$ runs {\em twice} over the interval $[0,\rho_{s}]$;
from $0$ to $\rho_{s}$ to cover the interior homogeneous
region, and then back from $\rho_{s}$ to $0$, to cover the
exterior Schwarzschild region\footnote{In the outside (Schwarzschild)
region, $\rho$ runs only over the semi-open interval $(0,\rho_{s}]$.
It does not reach $\rho=0$ which is asymptotic infinity $r=\infty$.}.

\vspace{0.4cm}

The quantization of this simple dust system is elementary.
It is easily done using (3) or (8). In coordinate
representation
  \be
\hat{X}(\rho,\eta) \equiv X(\rho,\eta) ~~~,~~~
\hat{P}_{X}(\rho,\eta) = -i \hbar \frac{\partial}{\partial X}
  \ee
we get the quantum version of Eq. (8), which is the
Schr\"{o}dinger equation
  \begin{eqnarray}
<X| \hat{H} |\Psi> &=&
\int \left( - \frac{\hbar^{2}}{2m(\rho)} \frac{\partial^{2}}{\partial
X^{2}} + \frac{1}{2}m(\rho) \omega^{2} X^{2} \right) \Psi[X;\eta]
d \rho \nonumber \\
&=& \int \tilde{E}(\rho) \Psi[X;\eta] d \rho
  \end{eqnarray}
where $ m(\rho) = \frac{3\pi}{\sqrt{1-\rho^{2}}} $ and $\omega = 1$.
We use a Schr\"{o}dinger quantization scheme, and not a
Wheeler-DeWitt one. Namely, the scalar product of quantum states
(which determines the Hilbert space) is the Schr\"{o}dinger
one\footnote{The Wheeler-DeWitt scalar product is of a Klein-Gordon
type [6,7]
  \begin{eqnarray*}
<\Psi_{1}|\Psi_{2}> = \int \psi_{1}^{*}(x)
\stackrel{\leftrightarrow}{\partial_{x}} \psi_{2}(x) ~.
  \end{eqnarray*} },
$<\Psi_{1}|\Psi_{2}> = \int \psi_{1}^{*}(x) \psi_{2}(x) dx$.
The solutions of (18) are harmonic oscillators, and the spectrum
$\tilde{E}(\rho)$ is quantized [3]. In the homogeneous case
this leads to a quantization condition for $a_{0}$ [6,8,9], and
in the collapsing OS case, it leads to a discrete mass [9,10].

The metric (6) is now an operator (replace $Y$ with the operator
$\hat{Y}$).
We would like to calculate the expectation value of this
quantized metric in a quasi-classical (coherent) state,
$|\alpha[X(\rho);\eta]>$.
We call the resultant matrix of real functions
  \be
g^{(\alpha)}_{\mu \nu}(\eta,\rho) \equiv <\alpha| \hat{g}_{\mu \nu} |\alpha>
  \ee
the ``quasi-classical metric", which corresponds to the classical
solution (9).
Using (18), we find the quasi-classical states,
which are the eigenstates of the annihilation operator
  \be
\hat{a}_{_{X}} |\alpha[X(\rho);\eta]> = \alpha[X(\rho);\eta] |
\alpha[X(\rho);\eta]>
  \ee
where
  \begin{eqnarray*}
\hat{a}_{_{X}} = \frac{1}{\sqrt{2}} \left( \sqrt{\frac{m(\rho)}{
\hbar} } \hat{X} + i \sqrt{ \frac{1}{\hbar m(\rho)}} \hat{P}
\right) ~~\mbox{and}~~
\alpha[X(\rho);\eta] = X_{0}(\rho) e^{i \eta}
  \end{eqnarray*}
and $X_{0}(\rho)$ is the classical initial value of $X$.
As in the standard harmonic oscillator case, the coherent
state, $|\alpha>$, is {\em not} and eigenstate of the
Hamiltonian (18). It is a superposition of such states.
But one can give the usual statistical interpretation,
it describes the probability of finding the system (collapsing
star or R-W universe) with a given mass (or energy).
The probability is described by a very sharp Gaussian
around the classical value.

Knowing the quasi-classical states (20), and the metric
operator which in the Heisenberg representation is
  \begin{eqnarray}
\hat{Y}(\eta,\rho) &=& X_{0}(\rho) \hat{1} + \hat{X}(\eta,\rho)
\nonumber \\
\hat{X}(\eta,\rho) &=& \sqrt{ \frac{\hbar}{2 m(\rho)} }
\left( \hat{a}_{x}(\rho) e^{-i\eta} + \hat{a}^{\dagger}_{x}(\rho)
e^{i\eta} \right) ~,
  \end{eqnarray}
we can calculate the quasi-classical metric.
One should notice however that we have an ambiguity
in the quantization scheme that we can use. The commutation
relations of $\hat{a}_{x}(\rho)$ and $\hat{a}^{\dagger}_{x}(\rho)$
can be functions of $\rho$. And different functions correspond
to different quantization schemes. This is related to the
ordering problem in our Schr\"{o}dinger equation [6,7].
At this point we don't need to determine the explicit
ordering that we use. We define the following creation
and annihilation operators
  \be
\hat{a} \equiv \frac{1}{\sqrt{m(\rho)} f(\rho)} \hat{a}(\rho)
{}~~~and~~~ \hat{a}^{\dagger} \equiv \frac{1}{\sqrt{m(\rho)} f(\rho)}
\hat{a}^{\dagger}(\rho)
  \ee
where the commutation relations of $\hat{a}$ and $\hat{a}^{\dagger}$
are $\rho$-independent, $[\hat{a},\hat{a}^{\dagger}] = 1$. The
$\rho$-dependent commutation relations of $\hat{a}_{x}(\rho)$ and
$\hat{a}^{\dagger}_{x}(\rho)$ is given by $m(\rho) f^{2}(\rho)$.
So the function $f(\rho)$ determine the quantization scheme
(the ordering) that we use.
With this definition we get the following metric expectation
value
  \begin{eqnarray}
<\alpha|d\hat{s}^{2}|\alpha> &=& (ds^{2})_{0} +
\frac{\hbar}{2} \left\{ - \frac{f^{2}}{\rho^{2}} d\eta^{2}
+ \left[ \left( (\partial_{\rho} f)^{2}
+ (\partial_{\rho} \eta)_{\tau}^{2} f^{2} \right) \frac{1}{1-\rho^{2}}
\right. \right. \nonumber \\
&-& \left. \left. \left( \partial_{\rho} \left( \frac{f}{\rho}
\right) \right)^{2} \eta^{2} \right] d\rho^{2}
- \left[ \partial_{\rho} \left( \frac{f^{2}}{\rho^{2}} \right)
\eta \right] d\eta d\rho + f^{2} d\Omega_{2}^{2} \right\}
  \end{eqnarray}
where $(ds^{2})_{0}$ is (6) with the classical $Y(\eta,\rho)$
solution (9). As expected, the quantum corrections are
proportional to $\hbar$, and depend on the quantization scheme
($f(\rho)$).

Consider first the homogeneous case. There is only one degree of
freedom in that case [2,9] , it is the scale factor, $a(\eta)$.
In that case $\Delta a$ and $\Delta P_{a}$ should {\em not}
be functions of $\rho$. This corresponds to choosing $f(\rho)=p\rho$
in (22), where $p$ is a constant\footnote{The uncertainty
principle is $(\Delta a)(\Delta P_{a}) = \frac{p^{2}}{2} \hbar$,
so $p^{2} \geq 1$.}. In that case we get [11]
  \be
<\alpha^{(h)}| d\hat{s}^{2} |\alpha^{(h)}> =
\left( a^{2}_{class}(\eta) + \frac{p^{2}}{2} \hbar \right)
\left( - d\eta^{2} + \frac{d\rho^{2}}{1
-\rho^{2}}+ \rho^{2} d\Omega_{2}^{2} \right)
  \ee
This describes a ball of dust that collapses to a minimum
radius (the Planck radius) and re-expands again [11]. The
quasi-classical metric (24) is singularity free.

In the Schwarzschild case the situation is more complicated but
as we are going to see, one can avoid the classical singularity
also in that case. Using the Schwarzschild radial
coordinate (14), we see that as long as $r^{2} >> \hbar f^{2}$,
the quasi-classical metric (23) is almost the classical
Schwarzschild one. The
deviations are of order $\hbar f^{2} / r^{2}$ or
$l_{P}^{2} f^{2}/r^{2}$. As we expected, the quantum
deviations are very small for $r > l_{P}$. On the other hand
near the classical singularity, $r=0$, the quantum corrections
are dominant. From (23) we have
  \begin{eqnarray}
\left( ds^{2}_{\alpha} \right)_{Sch} &=& - \left(\frac{\hbar}{2}
\frac{f^{2}}{\rho^{2}} + \mbox{o}(r^{2}) \right) d\eta^{2}
+ \left\{ \left( \frac{2Mr}{\rho^{2}} + \mbox{o}(r^{2}) \right)
\frac{ (\partial_{\rho} \eta)_{\tau}^{2} }{1-\rho^{2}} \right.
+ \nonumber \\
&+& \left. \frac{\hbar}{2} \left[ \left( (\partial_{\rho} f)^{2} +
(\partial_{\rho} \eta)_{\tau}^{2} f^{2} \right)\frac{1}{1-\rho^{2}}
- (\partial_{\rho} (f/\rho))^{2} \eta^{2} + \mbox{o}(r^{2})
\right] \right\} d\rho^{2} \nonumber \\
&-& \left( \frac{\hbar}{2} (\partial_{\rho} (f/\rho)^{2}) \eta +
\mbox{o}(r^{2}) \right) d\eta d\rho + \left( \frac{\hbar}{2}
f^{2} + \mbox{o}(r^{2}) \right) d\Omega_{2}^{2} ~,
  \end{eqnarray}
near $r=0$. We see that the dominant parts are the quantum
corrections. To study the structure of the metric (25)
(near $r=0$) we need to know $(\partial_{\rho} \eta)_{\tau}$.
One should remember that $\tau(\eta,\rho)$ is not the classical
function (12), there are quantum corrections. To find $\tau$ we
write the metric (23) in the form
  \be
ds^{2}_{\alpha} = -d\tau^{2} + g_{\tau \rho} d\tau d\rho
+ g_{\rho \rho} d\rho^{2} + \left( Y^{2}_{class} +
\frac{\hbar}{2}f^{2} \right) d\Omega_{2}^{2}
  \ee
and solve for $\tau(\eta,\rho)$. Generally it is quite
complicated to get $\tau$ explicitly, it is a complicated
function of the classical solution and $f(\rho)$. But near
$r=0$ one can easily solve for $\tau$ and get
  \be
\tau(\eta,\rho) \longrightarrow \sqrt{\frac{\hbar}{2}}
\eta \frac{f(\rho)}{\rho} + \tau_{0}
  \ee
and
  \be
(\partial_{\rho} \eta)_{\tau} \longrightarrow \eta
\partial_{\rho} \mbox{log}\left( \frac{\rho}{f(\rho)} \right)
  \ee
as $r \rightarrow 0$.

In the classical Schwarzschild case $(\partial_{\rho} \eta)_{\tau}$
diverge like $r^{-1}$ as $r$ goes to zero. This is
because of the classical singularity. In the quasi-classical
case we see (from (28)) that generally $(\partial_{\rho} \eta)_{\tau}$
is finite as $r$ goes to zero. From the
uncertainty principle we know that $f(\rho)$ cannot be
zero (in the interval $(0,\rho_{s}]$), so (28) cannot diverge
for a $C^{1}$-function, $f(\rho)$. If $f(\rho)$ is
also bounded, the Riemann tensor will be everywhere finite,
and generally its maximum will be around $r=0$ and of order
$M/l_{P}^{3}$. So for a general quantization scheme (for
which $f(\rho)$ is a bounded $C^{1}$ function), one {\em avoids}
the classical Schwarzschild singularity. Because there is
no singularity at $r=0$, one can extend the space-time
manifold to negative $r$'s.
Depending on the function $f(\rho)$, we may have different
situations: From (13) and (14)
we see that $r=0$ corresponds to $\eta = \pi$, so the
metric (25) on $r=0$ (and fix $(\theta , \phi)$), is
  \be
\left( ds_{\alpha}^{2} \right)_{r=0} = \frac{h}{2} \Gamma [f]
d \rho^{2}
  \ee
where
  \begin{eqnarray}
\Gamma [f] &=& \left( \frac{1+\pi^{2}}{1-\rho^{2}} \right)
(\partial_{\rho} f)^{2} - 2\pi^{2} \left( \frac{1}{\rho^{2}}
+ \frac{1}{\rho(1-\rho^{2})} \right) f (\partial_{\rho} f) +
\nonumber \\
&+& \pi^{2} \left( \frac{2}{\rho^{3}} +
\frac{1}{\rho^{2}(1-\rho^{2})} \right) f^{2}
  \end{eqnarray}
In this paper we examine only the cases in which $\Gamma [f]$
{\em does not} change sings in the interval $(0,\rho_{s}]$,
so we have three different situations: i) $\Gamma [f]
> 0$, in which case $r=0$ is spacelike, ii) $\Gamma [f] = 0$,
in which case $r=0$ is null, and iii) $\Gamma [f] < 0$, for
which $r=0$ is timelike. Typical Penrose diagrams of the
maximally extended manifolds for cases (i), (ii) and (iii) are
shown in Figures 1, 2 and 3, respectively.
In Fig. 1 the
manifold contains regions $(I)$ $r > r_{h}$, $(II)$ $0<r<r_{h}$,
$(II^{*})$ $-r_{h}^{*} < r < 0$, and $(I^{*})$ $r<-r_{h}$.
Where $r_{h} = 2M + \mbox{o}(M_{P}^{2}/M^{2})$, and
$-r_{h}^{*} = -2M + \mbox{o}(M_{P}^{2}/M^{2})$.
Regions $I^{*}$ and $II^{*}$ are very similar to $I$ and
$II$ respectively, but of course $r \rightarrow -r$
(as well as $t \rightarrow -t$) there.
In Fig. 2 the manifold contains the regions $(I)$ $r>r_{h}$,
$(II)$ $0<r<r_{h}$ and $(III)$ $r<0$. In this case, $r=0$ is
a Cauchy horizon like the inner horizon of a Kerr black hole.
In Fig. 3 the manifold contains the regions $(I)$ $r>r_{+}=r_{h}$,
$(II)$ $r_{-}<r<r_{+}$, $(III)$ $0<r<r_{-}$ and $(IV)$ $r<0$.
In this case, since $r=0$ is timelike there is also an inner
horizon ( at $ 0<r_{-}<r_{+}=r_{h}$ ) in which $ds^{2}_{\alpha}=0$.
This manifold is very similar to a Kerr one, but with no
singularity at $r=0$ (or anywhere else).

The manifolds in Figures 1, 2 and 3 are all complete. Namely,
in all of which all timelike curves have infinite proper length.
They all describe asymptotically flat universes
($r \rightarrow \pm \infty$) which are connected through
Planck size ``wormholes" ($r=0$).

{}From (14) we see that $r$ is the expectation value of
the dynamical field operator, $r = <\alpha| \hat{Y} |\alpha>$.
In a standard quantization one should consider both
possitive and negative field configurations, so the
quantum extension to negative $r$'s seems to be ``natural".

\vspace{0.4cm}

In this letter we use the midisuperspace of spherically
symmetric dust universes, suggested by Lund,
to study the quasi-classical
structure of the quantized Oppenheimer-Snyder collapsing
star. Especially, we find the outside quasi-classical
Schwarzschild metric. Depending on the quantization scheme
that we use, the classical singular hypersurface $r=0$,
becomes a {\em regular} spacelike, null or timelike hypersurface.
The quasi-classical metric is generally everywhere singularity
free. The maximal analytic extension of this space-time
is a complete manifold, with no singularities. It describes
a tower of asymptotically flat universes connected
through Planck size wormholes. The cases in which $r=0$ is
not always spacelike, null or timelike\footnote{This corresponds
to sings changes in $\Gamma [f]$ (see eq. (29)).}, require further
investigations.

Can we go beyond this midisuperspace model? Clearly, when we
include non-spherical deviations, we get propagating graviton
degrees of freedom, and the quantization become problematic.
It could be though, that non perturbative approaches may
give consistent results [12], otherwise we need to modify
Einstein gravity or quantum physics.
On the other hand, it may be interesting
to extend Lund's midisuperspace but still consider only
spherically symmetric geometries. For example, spherically
symmetric charged dust systems.

There is however one general feature that one can draw.
Our dynamical variable, $Y(\eta,\rho)$ is exactly the
vierbein, $e^{a}_{\mu}(x)$. In a general theory for  which
the vierbein is the basic dynamical variable, the metric
operator is its square, $\hat{g}_{\mu \nu} = \hat{e}^{a}_{\mu}
\hat{e}_{a \nu}$. If we use a free field representation for
the vierbein field, we can describe it as a collection
of harmonic oscillators, and have the standard quasi-classical
coherent state. The metric expectation value in this
quasi-classical state will be $<\hat{g}_{\mu \nu}>=(e^{a}_{\mu}
e_{a \nu})_{clas} + \mbox{o}(\hbar)$, and
the classical singularity, $e^{a}_{\mu} = 0$, can be avoided.

\vspace{0.7cm}
{\bf Acknowledgment} \\
I would like to thank Stanley Deser and Amos Ori for very
helpful discussions.

\newpage
{\bf REFERENCES}
\begin{enumerate}
\item R.C. Tolman, {\em Pric.Nat.Acad.Sci.}{\bf 20},
169 (1934) \\
L.D. Landau and L.M. Lifshitz, {\em The Classical Theory
of Fields}, (Pergamon, Oxford, 1962)
\item F. Lund, {\em Phys.Rev.}{\bf D8}, 3253; 4229 (1973)
\item Y. Peleg, ``Quantum Dust Black Holes", Brandeis Univ. Report No.
BRX-TH-350, hepth/9307057 (1993)
\item J.R. Oppenheimer and H. Snyder, {\em Phys. Rev.}
{\bf 56}, 455 (1939)
\item C.W. Misner, K.S. Thorne and J.A. Wheeler,
{\em Gravitation} (Freeman, San Francisco, 1973)
\item B. S. DeWitt, {\em Phys.Rev.}{\bf 160},1113 (1967)
\item J. A. Wheeler, in {\em Battelle Recontres}, eds. C.M.DeWitt
and J.A.Wheeler (Weily, New York, 1968)
\item J.H. Kung, ``Quantization of the Closed Minisuperspace
Models as Bound states", Harvard Univ. report, hepth/9302016
(1993)
\item Y. Peleg, ``The Wave Function of a Collapsing Star",
Brandeis Univ. Report No. BRX-TH-342, hepth/9303169 (1993)
\item J.D. Bekenstein, {\em Lett.Nuov.Cimen.}{\bf 11},
476 (1974) \\
V. F. Mukhanov, {\em JEPT Lett}. {\bf 44}, 63 (1986)\\
J.Garcia-Belido, ``Quantum Black Holes", Stanford Univ.
Report No. SU-ITP-93-4 (1993)
\item Y. Peleg, ``Avoidance of Classical Singularities in
Quantized Gravitational Dust Systems", Brandeis Report N0.
BRX-TH-354, hepth/940236 (1994)
\item A. Ashtekar, {\em Non-perturbative Canonical Gravity}
(World Scientific, Singapur, 1991).

\end{enumerate}

\newpage

{\bf Figure Captions}

\vspace{0.5cm}

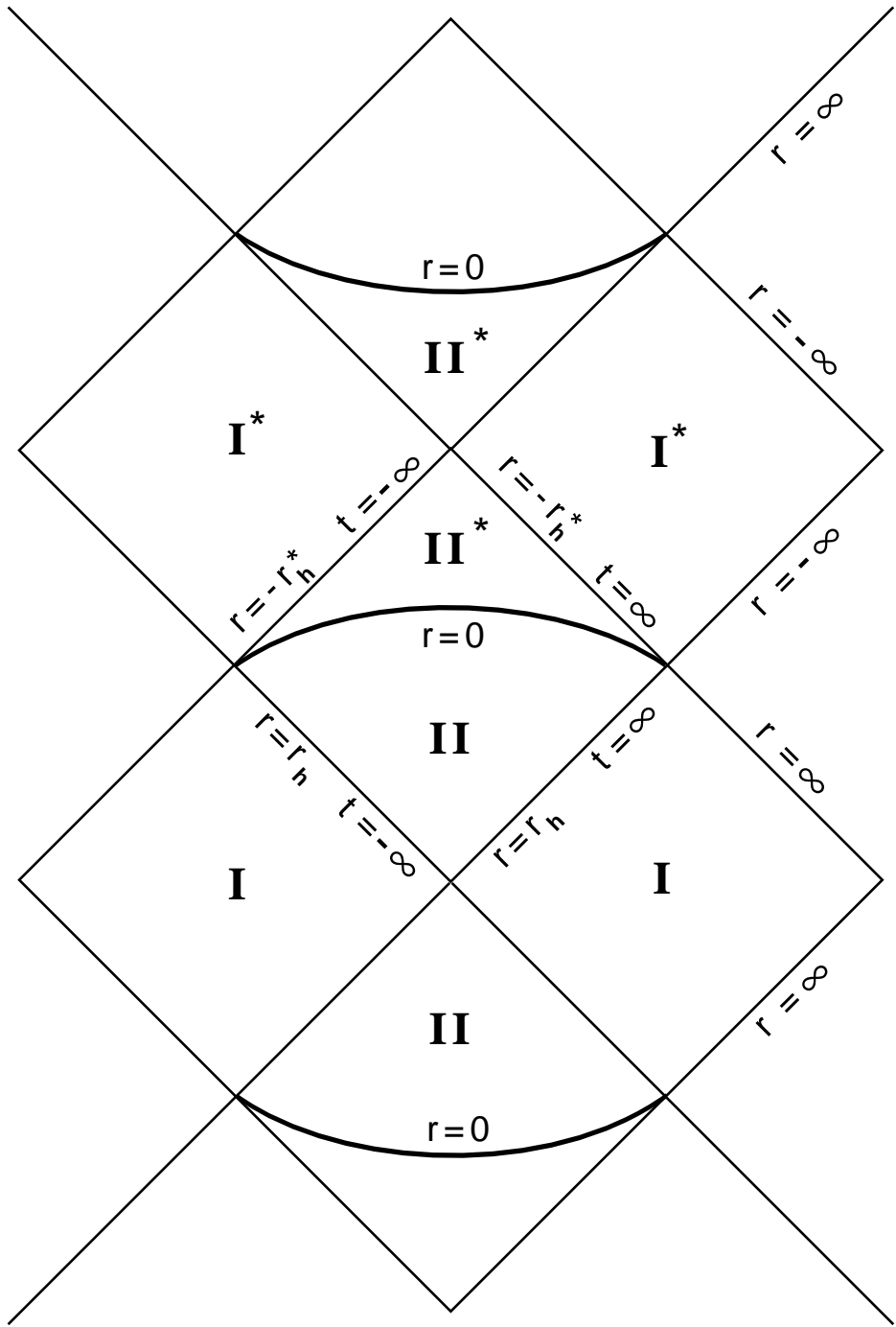
{\em Fig. 1:} Perose diagram for a typical maximally extended
manifold in the case where $r=0$ is spacelike.
This is an infinite tower of maximally extended
Schwarzschild spacetimes. Each two successive Schwarzschild
manifold, $M$ and $M^{*}$, are related by $r \rightarrow -r$
and $t \rightarrow -t$.

\vspace{1cm}

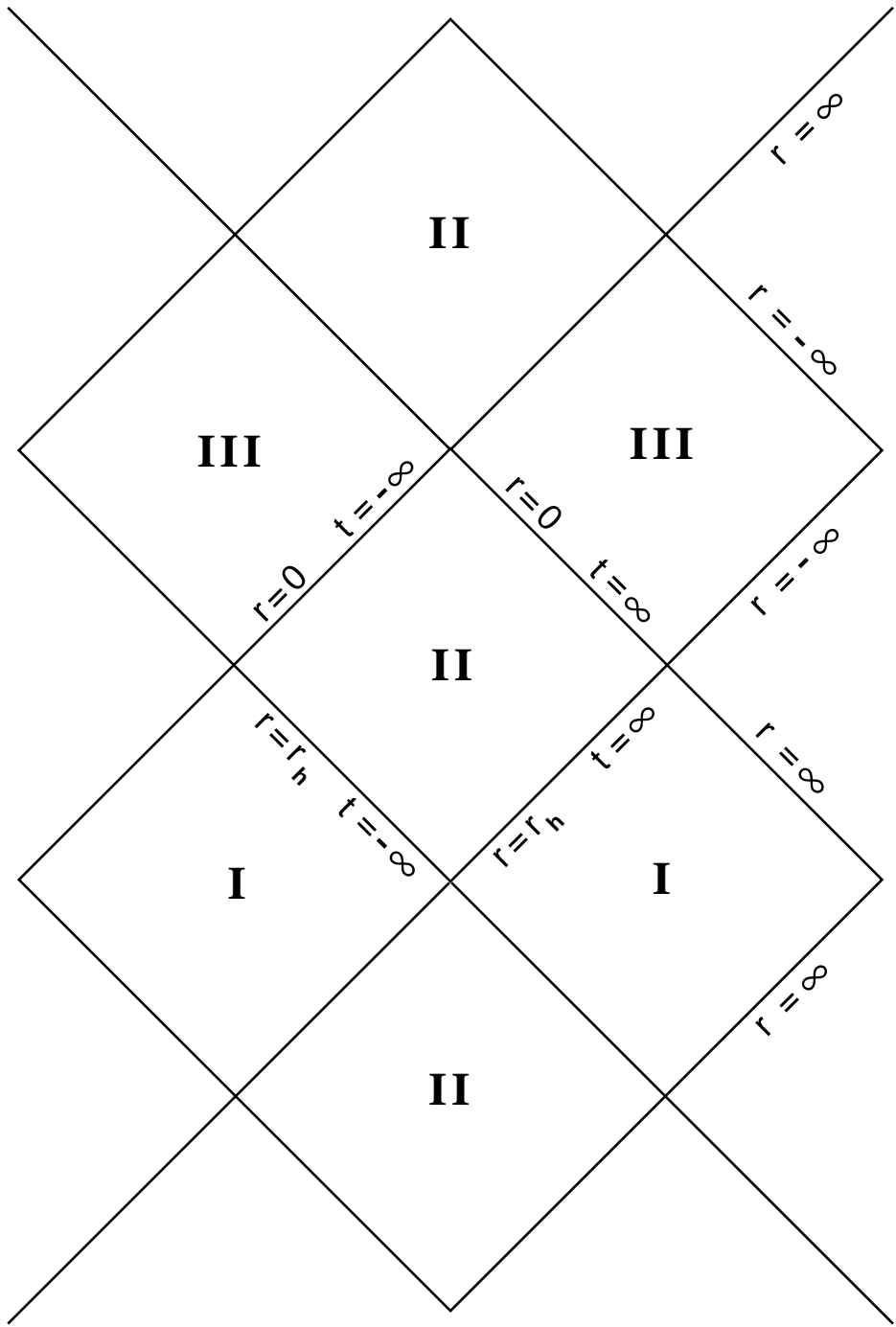
{\em Fig. 2:}  The Penrose diagram of the maximally extended
manifold in the case where $r=0$ is null. The classical singularity
$r=0$, becomes a Cauchy horizon.

\vspace{1cm}

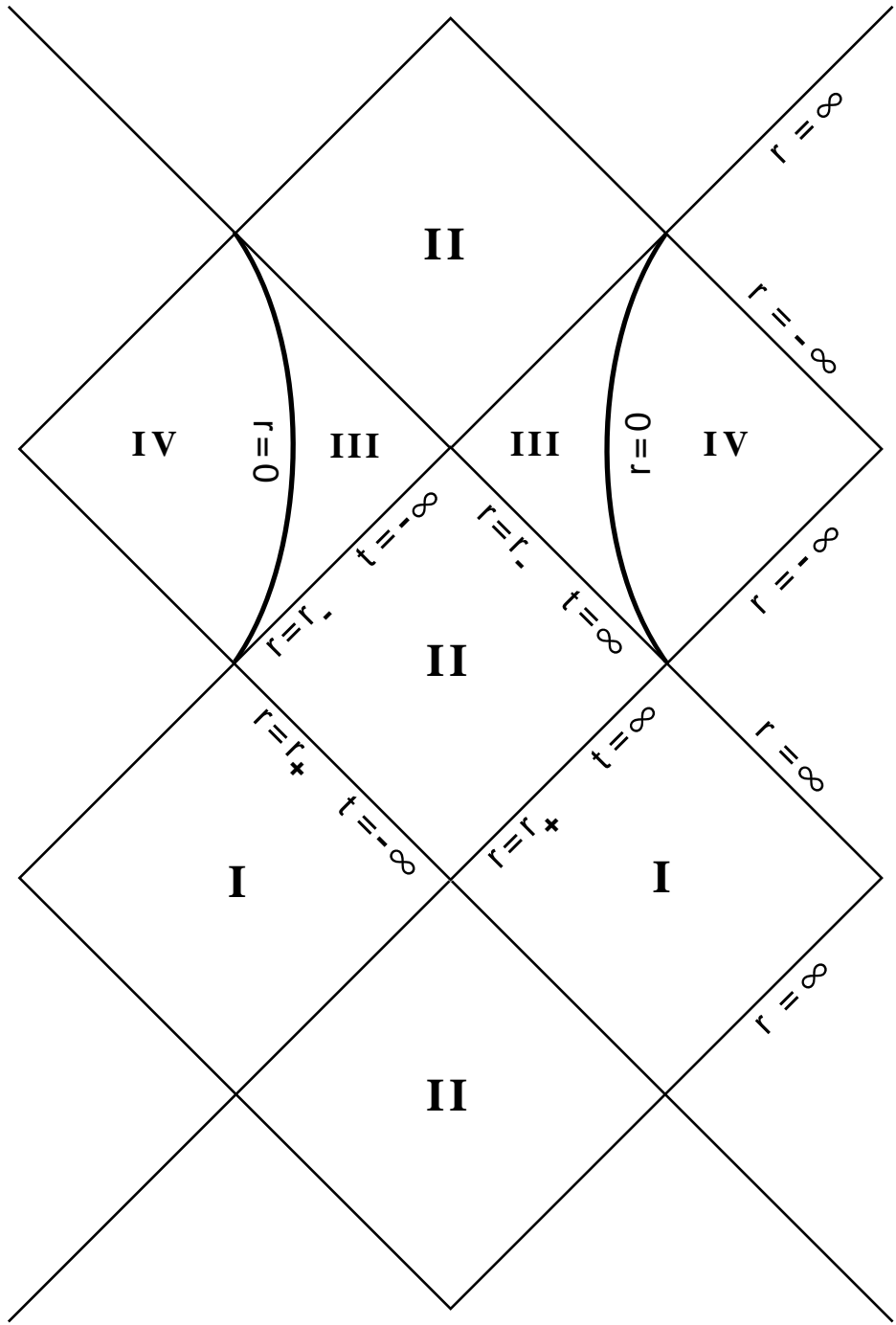
{\em Fig. 3:}  Penrose diagram for a typical maximally
extended manifold in the case where $r=0$ is timelike.
The manifold is very similar to
the maximally extended Kerr manifold, but there is no
singularity at $r=0$.

\end{document}